\documentclass[10pt,journal,compsoc]{IEEEtran}

\usepackage[normalem]{ulem}
\usepackage{amsmath, amsfonts, amsthm, amssymb, algorithm, algorithmicx}
\usepackage{algpseudocode}
\usepackage{changepage}
\usepackage{graphicx}

\usepackage{verbatim}

\usepackage{multirow}
\usepackage{longtable}
\usepackage{subfig}
\usepackage{url}

\ifCLASSOPTIONcompsoc
  \usepackage[nocompress]{cite}
\else
  \usepackage{cite}
\fi

\begin{document}

\title{
Cloud Server Benchmarks for Performance Evaluation of New Hardware Architecture}

\author{Hao Wu, Fangfei Liu, and Ruby B. Lee
\\  Princeton University, Princeton, NJ, 08544 USA
\\ {\footnotesize E-mail: haow.princeton@gmail.com, \{fangfeil, rblee\}@princeton.edu }
\IEEEcompsocitemizethanks{\IEEEcompsocthanksitem The authors are with the Princeton Architecture Laboratory for Multimedia and Security (PALMS), Department of Electrical Engineering, Princeton University, Princeton, NJ  08544\protect\\}

}

\IEEEtitleabstractindextext{%
\begin{abstract}
Adding new hardware features to a cloud computing server requires testing both the functionalities and the performance of the new hardware mechanisms. 
However,  commonly used cloud computing server workloads are not well-represented by the SPEC integer and floating-point benchmark and Parsec suites typically used by the computer architecture community.  Existing cloud benchmark suites for scale-out or scale-up computing are not representative of the most common cloud usage, and are very difficult to run on a cycle-accurate simulator that can accurately model new hardware, like gem5.  In this paper, we present PALMScloud,
a suite of cloud computing benchmarks for performance evaluation of cloud servers, that is ready to run on the gem5 cycle-accurate simulator.
We demonstrate how our  cloud computing benchmarks are used in evaluating the  cache performance of a new secure cache called Newcache as a case study.  
We hope that these cloud benchmarks, ready to run on a dual-machine gem5 simulator or on real machines, 
 can be useful to other researchers interested in improving hardware micro-architecture and cloud server performance.
\end{abstract}

}

\maketitle

\IEEEdisplaynontitleabstractindextext

%
\IEEEpeerreviewmaketitle

\ifCLASSOPTIONcompsoc
\IEEEraisesectionheading{\section{Introduction}\label{sec:introduction}}
\else
\section{Introduction}
\label{sec:introduction}
\fi

\IEEEPARstart Simulation of new hardware or new architecture is a necessary stage in computer hardware design, to test both its functionalities and its performance.
Ideally, we would like to evaluate the functionalities and performance on a detailed simulation platform at the design stage of the new systems before committing to expensive chip fabrication and prototype systems.  
There are open-source cycle-accurate simulation platforms that have been worked on for a long time (e.g., gem5\cite{gem5}, PTLsim\cite{Yourst2007}, MARSSx86 \cite{marss86}, etc), that simulate hardware at a very detailed level, and provide good performance data. They can also be leveraged to add new hardware features into the detailed hardware models.  

However, demonstrating performance of different hardware configurations requires representative benchmarks for the performance evaluation of today's computing environments.  
Frequently used benchmarks for performance may not be representative of current or future computing paradigms.  Currently, the computer architecture community uses SPEC Integer and Floating-point Benchmarks\cite{spec06} for general-purpose computing and some PARSEC benchmarks\cite{parsec} for parallel computing workloads.  These are mostly compute-intensive benchmarks and do not represent today's Cloud Computing scenarios, e.g., public Cloud Computing infrastructures like Amazon EC2\cite{amazon_ec2}.  The purpose of Cloud Computing  is to provide different IT resources (e.g. computing resources, storage resources, software development, system testing, etc. ) as services on demand to customers.  So the most critical attribute of Cloud Computing is resource virtualization.   

For example a company can buy real hardware resources (e.g. servers, etc.), or instead, buy Virtual Machines (VMs) from cloud infrastructure providers.   The company may deploy web servers and application servers in several of these VMs, and deploy data servers (e.g. MySQL) in other VMs.
In this paper, we did a study of the most common cloud computing services and workloads, and selected a suite of representative cloud computing benchmarks that can be run on a detailed hardware simulation platform.  With this paper, we will also make our Cloud benchmark suite, compiled to run on the gem5 simulator, available to the computer architecture research community.

To run these Cloud benchmarks under gem5, we had to get the x86 dual-computer simulation to work, to simulate both the server side and the client side of these benchmarks.  Then to demonstrate a concrete performance evaluation example, we use Newcache as a case study.  Newcache is a secure cache that uses dynamic randomized memory-to-cache mapping to prevent software cache side-channel attacks \cite{wang08, hasp13}.
By modifying the cache sub-system under gem5, we compared performance of different cache configurations.
The contributions of this paper are:
\begin{itemize}
\item A general-purpose and representative suite of cloud computing benchmarks, ready to run on gem5.
\item Open availability of the security simulation framework and cloud computing benchmarks, for computer architecture researchers.
\end{itemize}

Section \ref{sec:server-benchmarks} discusses representative cloud computing server benchmarks, and how they can be parameterized.
Section \ref{sec:dual_system} describes how we set up the x86 dual system in gem5.
Section \ref{newcache} describes our case study of a secure cache design, Newcache,
and gives sample performance testing results on Newcache used in place of conventional set-associative (SA) caches. 
Section \ref{sec:conclusions} concludes the paper.

\section{Cloud Computing Server Benchmarks}
\label{sec:server-benchmarks}

Cloud computing services can be categorized into several fundamental models\cite{book:cloud_comp}: Infrastructure as a service (IaaS), Platform as a service (PaaS), Software as a service (SaaS), etc.
Rackspace\cite{rackspace}, the leader in hybrid cloud and founder of OpenStack, put together a Top-10 list\cite{top10_uses_cloud} of the most common cloud computing use cases in 2012: File Storage and Sharing, Cloud Database, Email, PaaS for Web Applications, Web Site Hosting, etc.  
Some obsolete benchmarks (e.g., SPECweb\cite{spec}, SPECmail\cite{spec}, etc.) and currently in-use benchmarks (e.g., SPECjbb\cite{spec}, TPC-C\cite{tpc}, TPC-W\cite{tpc}, etc.) only cover some of these.  Also, since they target specific commercial purposes and include too many features, they run way too slow under a cycle-accurate simulator like gem5.  
CloudSuite\cite{cloudsuite} is a recent benchmark suite for scale-out workloads which covers many of the Top-10 list.  However, strictly speaking, most of them are big data applications (e.g. Hadoop Mapreduce for Data Analytics, Memcached for Data Caching, Cassandra NoSQL for Data Serving, etc. ), and they are not designed to work under a Cloud environment (virtual machines), so they are not really cloud server benchmarks.  
Virt-LM \cite{Huang2011} on the other hand is a suite of benchmarks used to evaluate the performance of live VM migration strategies among different software and hardware environments in a data center scenario, and Virt-LM chooses 5 popular cloud computing workloads, comprising 5 categories from Rackspace's Top-10 list.

Based on Virt-LM and Rackspace's Top-10 list, we carefully selected an initial set of 6 representative cloud computing workloads, including the workloads for a Web Server, Database Server, Mail Server, File Server, and Application Server, and also Streaming Server, which is not in Virt-LM or Rackspace's list, that are suitable to run on gem5.
Below, we describe possible programs for each server category, the programs we selected
, and the parameters we use for each program.  We also select a popular general purpose compute-intensive benchmark and an Idle Server benchmark for comparison.
Table 1 summarizes our server benchmarks and the client-side driving tools.


\begin{table}[htbp]
\scriptsize
  \centering
  \caption[Cloud Server Benchmarks in PALMScloud Suite]{Cloud Server Benchmarks in PALMScloud Suite}
  \label{tab:sum_cloud_benchmarks}
    \begin{tabular}{|l|l|l|}
    \hline
                       & \textbf{Server-side Benchmarks} & \textbf{Client-side Driving Tool} \\ \hline
    \textbf{Web Server}         & Apache httpd           & Apache ab                \\ \hline
    \textbf{Database Server}    & MySQL                  & SysBench                 \\ \hline
    \textbf{Mail Server}        & Postfix                   & Postal                   \\ \hline
    \textbf{File Server}        & Samba smbd             & DBench                   \\ \hline
    \textbf{Streaming Server}   & ffserver               & openRTSP                 \\ \hline
    \textbf{Application Server} & Tomcat                 & Apache ab                \\ \hline
    \textbf{Compute Server} & libsvm                 & a1a.t                \\ \hline
    \textbf{Idle Server} & -                 & -                \\ \hline
    \end{tabular}
\end{table}
\textbf{Web Server and Client: }
\textbf{Apache HTTP server (httpd)}\cite{apache_httpd} has been the most popular web server on the Internet since 1996.  The project aims to develop and maintain an open-source HTTP server for modern operating systems including UNIX and Windows NT.  
For the client side, we choose \textbf{Apache Benchmark Tool (ab)}\cite{apache_ab}, which is a single-threaded command line benchmarking tool well suited for a non-GUI testing environment under gem5.  This benchmark allows picking the number of total requests and the number of concurrent requests.  For example, to send 1000 HTTP requests to our Apache server with a concurrency of 10 requests at the same time, we type \emph{ab -n 1000 -c 10 http://10.0.0.1:8080/}.


\textbf{Database Server and Client: }
\textbf{MySQL}\cite{mysql} is a well-known open-source relational Database Management System (DBMS).  
To start MySQL server, we need to first create a user and group for the main program \emph{mysqld} to run.  Then we run script \emph{mysql\_install\_db} to set up initial grant tables, and finally we start the \emph{mysqld} service.  
For the client side, we choose \textbf{SysBench}\cite{sysbench}.  It is a modular, cross-platform and multi-threaded benchmark tool for evaluating OS parameters that are important for a system running a database under intensive load.  
In SysBench, we use the OLTP test mode, which benchmarks a real database's performance.  In the preparation phase, we create test tables with 100 records, while for the running phase, we do 200 advanced transactions.

\textbf{Mail Server and Client: }
We mainly focus on Simple Mail Transfer Protocol (SMTP) under gem5, and we choose \textbf{postfix}\cite{postfix} to act as the SMTP server.  We set the server side 10.0.0.1 to link with domain: domain1.com.  Postfix is on the server side, and is responsible for delivering all the emails.  The receivers of email are local users in domain1.com.
During the server initialization phase, we add 20 local users: user0 \url{~} user19 for our testing purposes.

For the client, we use \textbf{postal}\cite{postal} to send messages to the server.  \textbf{Postal} aims at benchmarking mail server performance.  It shows how fast the system can process incoming email.  We use command: \emph{postal -t X -m Y -c Z -f send-list 10.0.0.1 rcpt-list} to stress the server, where X is the number of threads attempting separate connections, Y is the maximum-message-size (in KB), and Z indicates messages-per-connection.  Also, send-list contains sender email addresses, and rcpt-list contains receiver email addresses (user0 \url{~} user19@domain1.com). In our experiment, we fix X=1, Y=1 and Z=3, and the stressing lasts 10 seconds (the time is controlled by linux terminal command wait()).


\textbf{File Server and Client: }
SMB provides file sharing and printing services to Windows clients as well as Linux clients.  We use the SMB protocol and choose \textbf{Samba smbd}\cite{smbd} as the file server in our test.  On the client side, we choose \textbf{Dbench}\cite{dbench} as the workload generator.  It can generate different I/O workloads to stress either a file system or a networked server.  We can first choose \textbf{dbench}'s  stressing backend (smb, nfs, iscsi, socketio, etc.) by specifying the -B option.  Since the server is  \textbf{Samba smbd}, we choose the backend to be smb.  Then we need to specify the shared file server folder and the user-password pair through the --smb-share option and the --smb-user option.  The shared folder and the user-password pair are already set up by the \textbf{smbd} server.  However in our experiments, we don't use a user-password pair.  Moreover, \textbf{dbench} has a key concept of a ``loadfile'', which is a sequence of operations to be performed on the file server's shared folder.  The operations could be ``Open file 1.txt'', ``Read XX bytes from offset XX in file 2.txt'', ``Close the file'', etc.  In our experiments, we generate two different ``loadfiles'',  one is a write-intensive load (smb-writefiles.txt), another is a read-intensive load (smb-readfiles.txt).  Finally, we can add a number \textbf{n} at the end of the \textbf{dbench} command to specify the total clients simultaneouly performing the load.

In the results shown in section \ref{newcache},  we type \emph{./dbench -B smb --smb-share=//10.0.0.1/share --smb-user=\% --loadfile=smb-writefiles.txt --run-once --skip-cleanup 3} to generate smbd.write, which means launching 3 clients (simulated as processes), and each client opens and writes five 64kB files (the sequence of operations are specified in loadfile=smb-writefiles.txt).  By replacing loadfile with smb-readfiles.txt, we generate smbd.read, which launches 3 clients and each client opens and reads five 64kB files.

\textbf{Streaming Server and Client: }
We use \textbf{ffserver}\cite{ffserver} as our streaming server.  It is a streaming server for both audio and video, which can stream mp3, mpg, wav, etc.  ffserver is part of the ffmpeg package.  It is small and robust.  Before starting the server, we need to register server side media-files at ffserver.conf file.  
On the streaming client side, we choose \textbf{openRTSP}\cite{openrtsp}.
RTSP protocol can basically control the streaming.  Clients issue VCR-like commands, such as play and pause, to facilitate real-time control of playback of media files from the server.   In the experiment, ffserver registered several mp3 audio files on the server side, and it uses port 7654 for streaming.
On the client side, we use command: \emph{./openRTSP -r -p [port-number] rtsp://10.0.0.1:7654/*.mp3} to generate a streaming request, where `-r' indicates playing the RTP streams without receiving them, and `-p' option indicates the local port to stream the file.
In section \ref{newcache}, ffserver.sX means using \textbf{openRTSP} to send X different remote client connection requests to the ffserver for mp3 files streaming.  We generate 3 different streaming workloads (X=1, X=3 and X=30) to test the streaming server.  

\textbf{Application Server and Client: }
For web applications, the application server components' main job is to support the construction of dynamic pages\cite{app_server}.  
Application servers differ from web servers by dynamically generating html pages each time a request is received, while most http servers just fetch static web pages.  Application servers can utilize server-side scripting languages (PHP, ASP, JSP, etc.) and Servlets to generate dynamic content.
We use \textbf{Tomcat}\cite{tomcat} on the server side.  
It is a small, robust application server that also provides us with a lot of useful small jsp and servlet examples. 
For the testing, we use \textbf{Apache ab}\cite{apache_ab} to send HTTP requests to \textbf{Tomcat} (Apache ab is described previously as a web server client).  
We use command: \emph{ab -n Y -c Z http://10.0.0.1:8080/(X URLs)}, where X represents X different URLs, Y is the number of requests to perform for each URL, and Z is the requesting concurrency for each URL.  These different URLs contain different jsp and servlet examples provided by \textbf{Tomcat}.  
In our experiments, we fix Y=10 and Z=2, and choose X to be 1, 3 and 11, respectively, for three different workloads, ranging from light to heavy work.

\textbf{Computation Machine Learning Server: } 
We also want to use some computing benchmarks to represent compute servers.  We choose LIBSVM \cite{CC01a} to do Support Vector Machine (SVM) classification.  On the server side, we have already trained an SVM model using data set a1a's training data from UCI's machine learning repository \cite{Lichman:2013}. a1a is called Aduilt Data Set, which can be used to predict whether a person's income exceeds \$50K/yr based on his census data.  We use this model to do SVM classification on a1a's testing data: a1a.t.   The testing data can either be placed on the server side in advance, or can be transfered from the client side (using netcat, php file upload, file server, ssh, etc.) during runtime.  

\textbf{Idle Server and Client: } 
We run a clean server without any client-side stressing as our baseline benchmark.

\section{Simulating Client-Server Dual System in Gem5}
\label{sec:dual_system}
The original gem5 simulator had a python script for configuring the dual system for ALPHA, ARM and MIPS, but didn't provide a python script for setting up the dual system for x86, due to the complexity of the x86 system. The python script in gem5 is used to define the connection of the system and will be used to automatically generate C++ code. Basically, to set up a dual system, we need to add an Ethernet interface to each system and connect them using an Ethernet link. The Ethernet interface is a PCI device and will be connected to the Intel chip-set's North-Bridge chip. The configuration not only specifies how the Ethernet device is connected to the system, but also the proper assignment of the interrupt number which is defined in the MP Table (Multiprocessor Configuration Table), as well as the address range of the MMIO (Memory-Mapped I/O). In addition to the python script, we need to patch the gem5 source code so that the cpuid can support long address mode and the access to the Ethernet device is uncacheable.

\begin{figure}[htbp]
  \centering
    \includegraphics[scale=0.3]{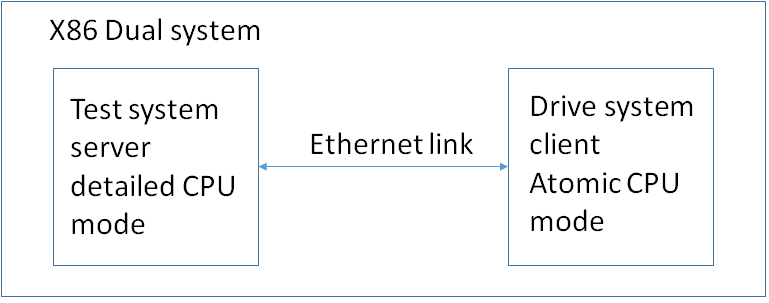}
  \caption[New x86 Dual System in gem5]{New x86 Dual System in gem5}
  \label{fig:x86_dual_system}
\end{figure}

In Figure \ref{fig:x86_dual_system}, after these x86 dual-system configuration settings under gem5,  we can simulate both server side (Test system) and client side (Drive system), and they are able to communicate via the Ethernet link.  The server has an IP address 10.0.0.1, and the client has an IP address 10.0.0.2.   By taking advantage of the dual system, we can adjust the driving workload from the client side and measure the performance on the server side, instead of using a server-side fixed input script.  Also, the dual system simulates a real two-node network environment.  The server applications provide services and monitor input requests on their specified TCP ports, while the client programs request for services (HTTP requests, streaming requests, etc.) through these server-side ports.


Both server side and client side use the basic Linux system installed on gem5's image disk.  The server has different server-side applications installed. When booting the dual system, both server and client configure their network interface.  The server then initiates the server-side services and sends a ‘ready’ signal to the client, so that the client can drive the server side with requests.
Some Server-side benchmarks actually require some time to start up before their services are available to the client under gem5.  We use a daemon program to test periodically if the service ports are opened up by the benchmark.  Only at that time will the server send a 'ready' signal to the client.

\begin{figure*} [htbp]
\centerline{ \subfloat [] {\includegraphics[trim=2.5cm 9.2cm 1cm 9cm, width=3.8in]{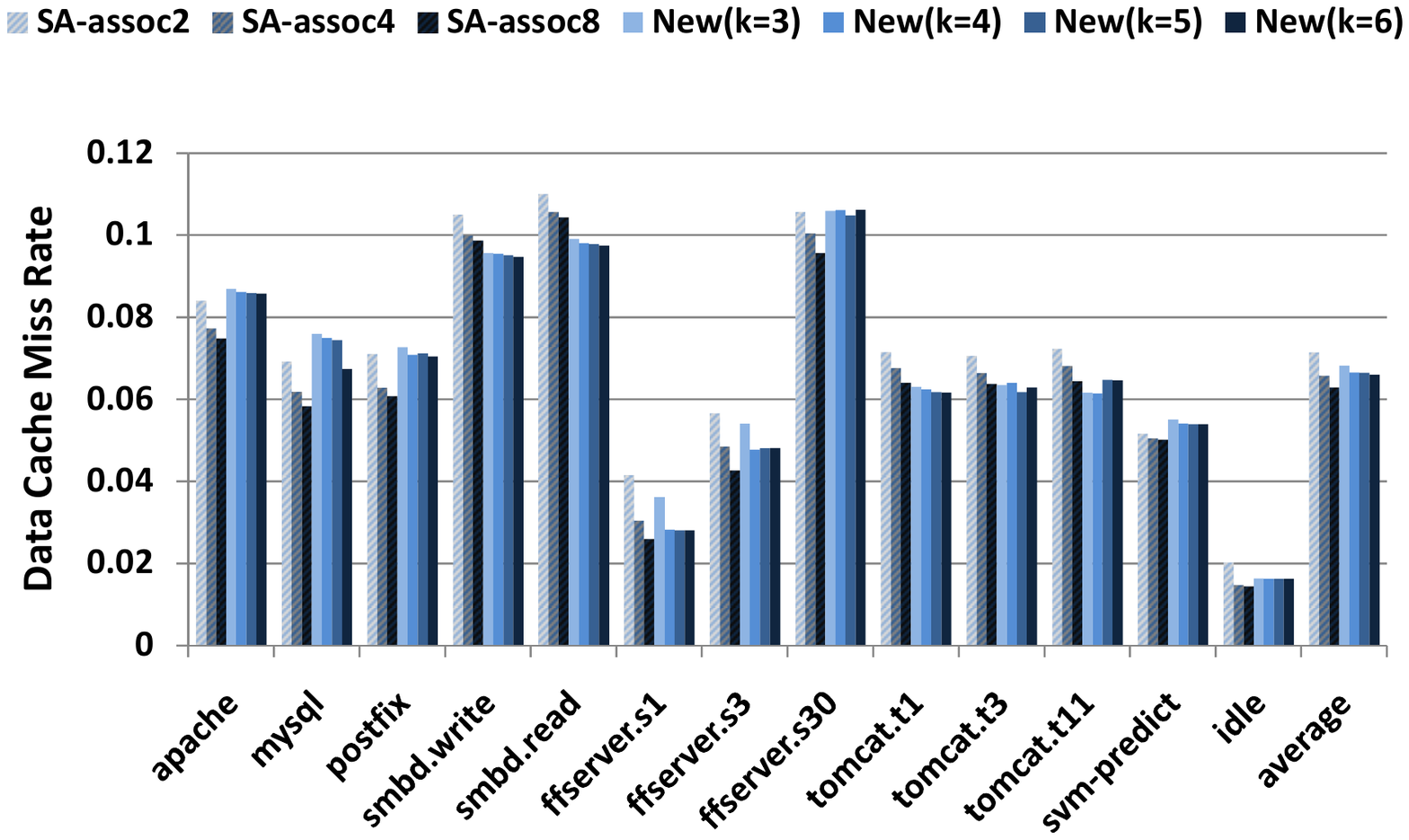}
\label{fig:l1d_dcache-missrate_2}}
\hfil
\subfloat[] {\includegraphics[trim=2.5cm 9.2cm 1cm 9cm,width=3.8in] {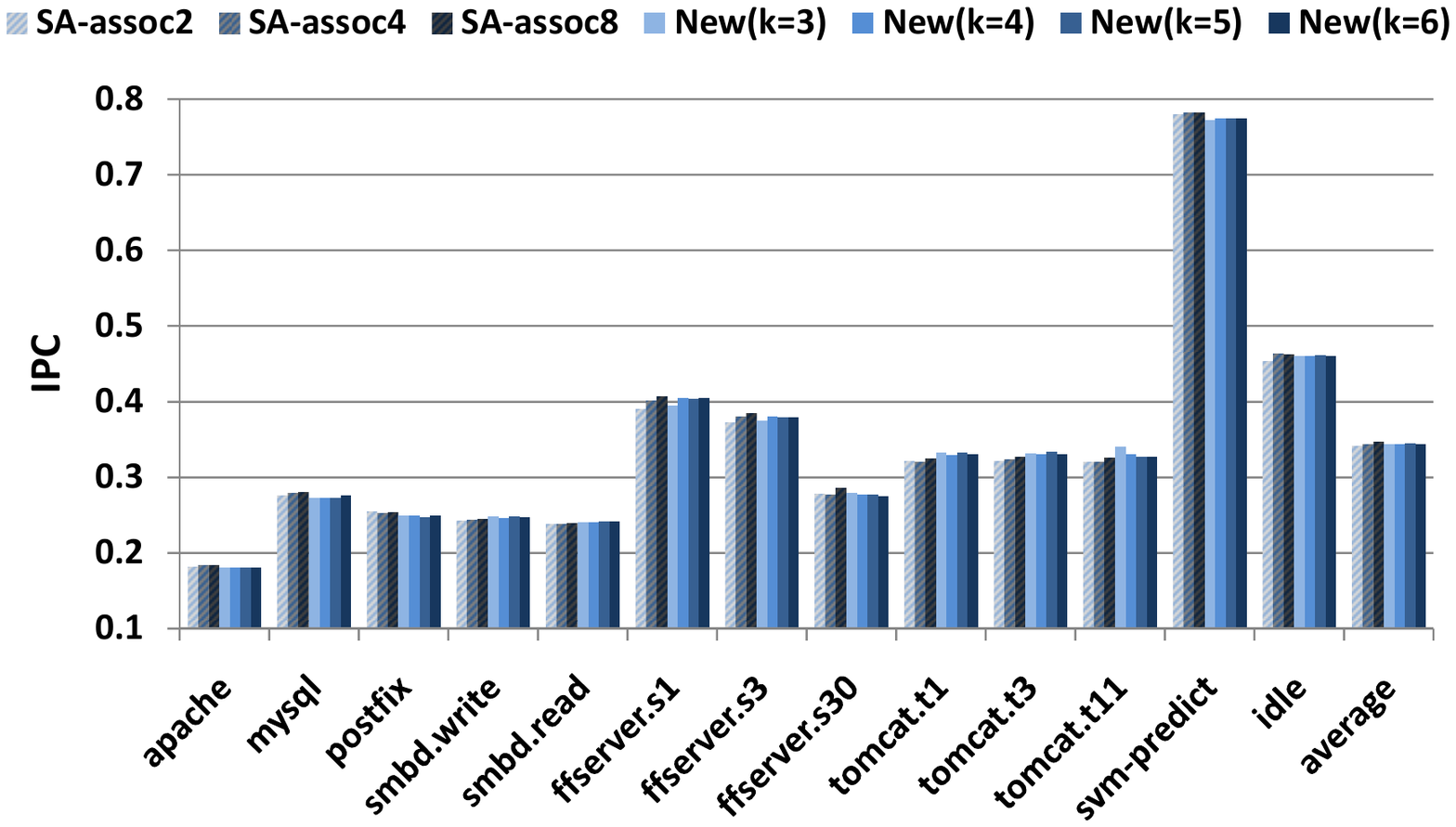}
\label{fig:l1d_ipc_2}}}
\caption{Newcache as L1 D-cache: (a) Data Cache Miss Rate, and (b) Instructions Per Cycle (IPC) for Different Associativity and Newcache extra index bits, k.  Here, the cache size is fixed at 32kB}
\label{fig:l1d_newcache}
\end{figure*}


\section{Example: Secure Caches}
\label{newcache}

Recently, much interest in the security world focused on using "Moving Target Defense" (MTD) to thwart the attacker, by having the system appear to change continuously. We notice with interest that Newcache \cite{wang08}, a proposal for a secure cache that thwarts cache side-channel attacks, is an example of using MTD in hardware design.  However, the performance of Newcache was only tested with SPEC benchmarks\cite{wang08}.  Hence, we use Newcache as a concrete example for performance evaluation with current Cloud Computing server benchmarks (section \ref{sec:server-benchmarks}).  You can refer to \cite{hasp13, hasp15} to see both security testing and performance testing of Newcache.  We model Newcache in a modular way, so that it can replace any of the caches in the cache hierarchy in gem5.


\begin{table}[htbp]

  \scriptsize
  \centering
  \caption[Gem5 Base CPU and Cache Configurations]{Gem5 Base CPU and Cache Configurations}
  \label{tab:gem5_base}
\begin{tabular}{|l|l|}
\hline                        
   \textbf{Parameter} & \textbf{Value} \\
  \hline 
   L1 data cache (private) associativity, and size & 8-way SA, 32 KB \\ [0.5ex]
    \hline 
   L1 instruction cache (private) associativity, and size & 4-way SA, 32 KB \\[0.5ex]
   \hline
  L2 cache (private) associativity, and size & 8-way, 256 kB \\ [0.5ex]
    \hline 
  L3 cache (shared) associativity, and size & 16-way, 2MB \\ [0.5ex]
    \hline
  Cache line size & 64 B \\ [0.5ex]
    \hline 
  L1 hit latency & 4 cycles  \\ [0.5ex]
    \hline
  L2 hit latency & 10 cycles \\[0.5ex]
    \hline
  L3 hit latency & 35 cycles \\[0.5ex]
    \hline
  Memory size, and latency & 2 GB, 200 cycles \\ [0.5ex]
    \hline
\end{tabular}

\end{table}

\noindent\textbf{Newcache as L1 Data Cache: }
 We replace the L1 data cache with Newcache, and compare the performance results against a baseline x86 system with Set-Associative (SA) caches, with parameters shown in Table \ref{tab:gem5_base}, like those of the Intel core i7.  
We run the benchmark services on the server side and driving tools on the client side.  The results for Instructions Per Cycle (IPC) and Data Cache Miss Rate for all the benchmarks are shown in Figure \ref{fig:l1d_newcache}.  In the figure, k is the number of extra index bits for Newcache to have better security and performance.  
We also study many other different configuration parameter values.  For our case study of Newcache, we found that it does not degrade overall IPC performance when used as a D-cache, for our cloud server benchmarks, when compared to conventional SA caches.

\section{Conclusions}
\label{sec:conclusions}
We have defined a new Cloud benchmark suite that represents the common workloads of today's public cloud computing usage, and we described how to run such client-server benchmarks in x86 dual-system mode on gem5.   We hope that this paper and the Cloud benchmark suite under the enhanced gem5 simulator can expedite cloud performance evaluation for other researchers when new hardware features are simulated on gem5.  Since security testing can also be done on gem5 \cite{hasp13, hasp15}, we also hope to enable better security-performance tradeoff studies, and thereby speed up the design of better hardware-software security mechanisms and architectures.

\bibliographystyle{IEEEtranS}
\bibliography{references}

\end{document}